# Bayesian Watermark Attacks

**Ivo D. Shterev**                                                                                     I.SHTEREV@DUKE.EDU
Duke University, Durham NC 27705, USA

**David B. Dunson**                                                                                    DUNSON@STAT.DUKE.EDU
Duke University, Durham NC 27708, USA

## Abstract

This paper presents an application of statistical machine learning to the field of watermarking. We propose a new attack model on additive spread-spectrum watermarking systems. The proposed attack is based on Bayesian statistics. We consider the scenario in which a watermark signal is repeatedly embedded in specific, possibly chosen based on a secret message bitstream, segments (signals) of the host data. The host signal can represent a patch of pixels from an image or a video frame. We propose a probabilistic model that infers the embedded message bitstream and watermark signal, directly from the watermarked data, without access to the decoder. We develop an efficient Markov chain Monte Carlo sampler for updating the model parameters from their conjugate full conditional posteriors. We also provide a variational Bayesian solution, which further increases the convergence speed of the algorithm. Experiments with synthetic and real image signals demonstrate that the attack model is able to correctly infer a large part of the message bitstream and obtain a very accurate estimate of the watermark signal.

## 1. Introduction

Watermarking is the process of imperceptibly embedding a watermark signal into a host signal (audio segment, pixel patch from image or video frame). The watermark signal should only introduce tolerable distortion to the host signal and it should be recoverable by the intended receiver. Watermarking techniques differ by the way they modulate the host signal to embed information. There are two major classes of watermark embedding schemes, namely spread spectrum and quantization index modulation (QIM) (Chen & Wornell, 2001).

Spread spectrum watermarking (Cox et al., 2007; Hartung et al., 1999) constitutes a popular class of watermarking algorithms. In their simplest form, the watermarked signal is constructed by adding the host and watermark signals together, i.e. additive watermark embedding. Although, in terms of additive noise attacks they have been outperformed by the more robust QIM watermarking techniques (Chen & Wornell, 2001), spread-spectrum techniques have advantageous features that make them preferable in some watermarking scenarios. Examples of such inherent features include their simplicity and robustness to removal attacks. Another advantage is that spread-spectrum watermarking can be applied in different forms (multiplicative watermarking (Huang & Zhang, 2007)) that can further improve performance in some cases. They can also effectively exploit the human visual system (HVS) (Podilchuk & Zheng, 1998) to reduce perceptual degradation of the host signal.

Many attacks have been designed to hamper the performance of watermarking in general and spread-spectrum watermarking in particular. The attacks are usually classified with respect to the attacker's assumed knowledge about the watermark scheme. Robustness attacks pertain to the class of attacks under which the attacker has no knowledge of the watermark scheme. Examples of such attacks include adding random noise (Chen & Wornell, 2001) to the watermarked signal, replacing signal blocks with perceptually similar blocks computed in a certain way (Kirovski et al., 2007), applying a geometric transformation (cropping, scaling, translation, etc.) to the watermarked signal, or applying a malicious filtering operation (Su et al., 2001), to name a few. Other attacks belong to the so called worst case class of attacks, where the attacker has knowledge about the watermark technique and designs the attack such that the watermark detector (decoder) performance is minimized, under suitably defined distortion constraints. Usually, this type of attack is based on game theory (Cohen & Lapidoth, 2002) and is mostly of theoretical importance.

A third class of attacks aims at compromising the watermark system security (security attacks) (Cayre et al., 2005; Freire & Gonzalez, 2009). Under this scenario, the attacker has access only to the watermarked data and tries to estimate the secret key used for embedding the watermark. Having estimated the secret key, he can then reconstruct the watermark and remove it from the watermarked data (the so called removal attacks), thus creating a forgery of the host signal, which can then be freely copied and distributed by pirates. Although (Cayre et al., 2005; Freire & Gonzalez, 2009) develop theoretical security attack frameworks, the proposed algorithms do not perform well with real correlated host signals.

Another type of attack is the so called sensitivity analysis attack, which constitutes a powerful subclass of removal attacks (Kalker et al., 1998; Linnartz & van Dijk, 1998; Choubassi & Moulin, 2007). In their attempt to estimate the watermark signal, they rely on unlimited access to the decoder.

In this paper, we consider the scenario in which a watermark signal is repeatedly embedded in specific (possibly secretly chosen) host signals. The host signal can represent a patch of pixels from image or video frame. The host signals may be perceptually similar or quite disparate, as the watermark algorithm may choose, for security reasons, to embed the watermark in specific signals of the host data based on a secret message bitstream. Repetitive watermark embedding is of particular interest in image and video watermarking (Voloshynovskiy et al., 2001; Lu & Hsu, 2007; Bas et al., 2002; Tang & Hang, 2003; Doerr & Dugelay, 2004; Kalker et al., 1999), where the watermark signal is repeatedly allocated into small blocks to ensure robustness and resistance to geometric (desynchronization) attacks. However, the proposed attacks related to this scenario assume that the watermark signal is not secretly hidden but is added to every host signal and therefore do not try to estimate an embedded message bitstream.

The attack model proposed in this paper jointly estimates the embedded message bitstream and watermark signal from the watermarked data, without access to the decoder. We develop a probabilistic model based on Bayesian statistics. The algorithm models the host signal as having a multivariate Gaussian distribution with unknown mean and full covariance matrix. The watermark signal itself is also modeled as having a multivariate Gaussian distribution, but with separate unknown mean and full covariance matrix. The model parameters are updated sequentially from their respective conjugate full conditional posterior distributions, via Markov chain Monte Carlo (MCMC) sampling. To further increase the convergence speed of the proposed algorithm, we develop a variational Bayesian (VB) (Beal & Ghahramani, 2003) solution to it. In addition to its suitability for large scale data analysis, the VB solution also allows for diagnosing convergence, via the lower bound to the log-likelihood. Both MCMC and VB solutions perform comparably with respect to probability of bit error and relative watermark reconstruction error, with both synthetic and real host data.

Our model borrows similar ideas from sparse factor regression formulations (sparse models) used in gene expression data analysis (West, 2003; Carvalho et al., 2008). The objective of such sparse models is to specify a prior for the elements of a highly sparse factor loadings matrix, with most elements being exactly zero and few of them having relatively large variances. To contrast with our model, the role of the zero elements here is taken by the data points (signals) that are not watermarked, which do not necessarily constitute the majority of all data points. The watermarked data points have the interpretation of the non-zero elements in the factor loadings matrix. However in our case, they are a sum of the host and the watermark signals, with the watermark signal being much *weaker* than the host signal. The problem becomes that of a joint identification-estimation of a *subtle* signal.

## 2. Spread-Spectrum Watermarking

Throughout this paper, random variables are denoted by small letters. Random vectors and their realizations are denoted by bold small letters. The notation $\mathbf{x} \in \mathcal{R}^d$ indicates a $d$-dimensional random vector of real elements. Square random matrices and their realizations are denoted by bold capital letters. The notation $\mathbf{X} \in \mathcal{R}^{d \times d}$ indicates a $d \times d$ matrix of real elements and $\mathbf{X}'$ is its transpose. The probability of an event is denoted by $\Pr(\cdot)$. The notation $x \sim p(x)$ indicates that $x$ has a probability density function (pdf) p(x).

In this paper we concentrate on one of the most popular additive spread-spectrum watermarking systems, in which a watermark signal is repeatedly used to embed a message bitstream into a host data. The watermark encoder is shown in Fig. 1. Considering

Bayesian Watermark Attacksthe $i$th data point and depending on the message bit $b_i \in \{0,1\}$, the encoder adds ($b_i = 1$) the watermark signal $\mathbf{w}$ to the host signal $\mathbf{x}_i$, or leaves the host signal unchanged ($b_i = 0$). The watermarked signal can therefore be written as

$$\mathbf{y}_i = \begin{cases} \mathbf{x}_i + \mathbf{w} & \text{if } b_i = 1, \\ \mathbf{x}_i & \text{if } b_i = 0, \end{cases} \quad (1)$$

where $i \in \{1,\ldots,n\}$ and $n$ is the number of available data points.

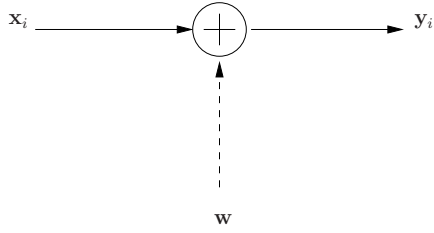

Figure 1. Additive spread-spectrum watermark encoder.

The watermark decoder is shown in Fig. 2. The decoder has access to the watermark signal $\mathbf{w}$. Based on the received (watermarked) signal $\mathbf{y}_i$ and the watermark signal, the decoder computes a detection test statistic $f(\mathbf{y}_i, \mathbf{w})$ and compares it to a suitably chosen threshold $\tau$. The decoder then outputs an estimate $\hat{b}_i$ of the embedded message bit $b_i$ in the following way

$$\hat{b}_i = \begin{cases} 1 & \text{if } f(\mathbf{y}_i, \mathbf{w}) > \tau, \\ 0 & \text{if otherwise.} \end{cases} \quad (2)$$

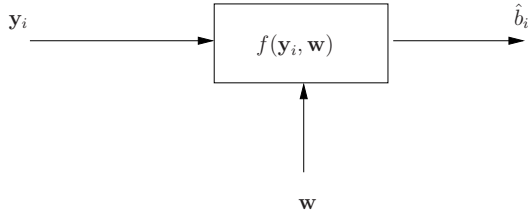

Figure 2. Watermark decoder.

Throughout the paper, the document-to-watermark (DWR) ratio is defined as DWR $= 10\log_{10} \sigma_x^2/\sigma_w^2$, where $\sigma_x^2$ is the variance of a single element in $\mathbf{x}_i$, for $i \in \{1,\ldots,n\}$, and $\sigma_w^2$ is the variance of a single element in $\mathbf{w}$.

## 3. Attack Model

It is assumed that the attacker has access to the watermarked signal, but has no access to the watermark decoder. The complete form of the attack model can be summarized as follows:

$$\begin{align}
\mathbf{y}_i &= \mathbf{x}_i + b_i \mathbf{w} & (3) \\
\mathbf{x}_i &\sim \mathcal{N}(\mathbf{x}_i|\boldsymbol{\mu}, \boldsymbol{\Sigma}) & (4) \\
\mathbf{w} &\sim \mathcal{N}(\mathbf{w}|\mathbf{m}, \mathbf{V}) & (5) \\
b_i &\sim \text{Bernoulli}(b_i|\pi) & (6) \\
\{\boldsymbol{\mu}, \boldsymbol{\Sigma}\} &\sim \mathcal{N}(\boldsymbol{\mu}|\boldsymbol{\mu}_0, \boldsymbol{\Sigma})\mathcal{IW}(\boldsymbol{\Sigma}|\omega_0, \boldsymbol{\Sigma}_0) & (7) \\
\{\mathbf{m}, \mathbf{V}\} &\sim \mathcal{N}(\mathbf{m}|\mathbf{m}_0, \mathbf{V})\mathcal{IW}(\mathbf{V}|\omega_0, \mathbf{V}_0) & (8) \\
\pi &\sim \text{Beta}(\pi|a_\pi, b_\pi), & (9)
\end{align}$$

where $\mathcal{N}(\mathbf{x}|\boldsymbol{\mu}, \boldsymbol{\Sigma})$ is the $d$-variate Gaussian distribution of $\mathbf{x}$ with mean $\boldsymbol{\mu}$ and covariance matrix $\boldsymbol{\Sigma}$, $\mathcal{IW}(\boldsymbol{\Sigma}|\omega_0, \boldsymbol{\Sigma}_0)$ is the inverse Wishart distribution of $\boldsymbol{\Sigma}$ with degrees of freedom $\omega_0$ and base covariance matrix $\boldsymbol{\Sigma}_0$, Bernoulli($b_i|\pi$) is the Bernoulli distribution of $b_i$ with mean $\pi$, and Beta($\pi|a_\pi, b_\pi$) is the Beta distribution of $\pi$ with parameters $a_\pi$ and $b_\pi$.

A graphical representation of the attack model is shown in Fig. 3. The blue circle represents the observed variable, the white circles represent hidden (latent) variables and the squares represent hyperparameters. Conditional dependence between variables is shown via the directed edges.

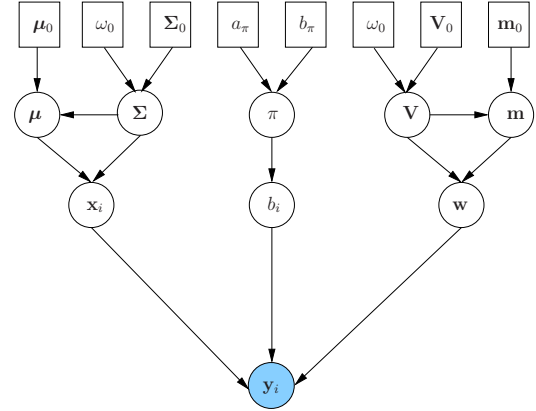

Figure 3. A graphical representation of attack model.

## 4. Posterior Updates

In this section we derive the update equations for the attack model parameters, with respect to the MCMC and VB solutions. The update equations are based on the full likelihood of the model, which can be written as

$$\begin{align}
\mathcal{L}(\mathbf{y}) &= p(\mathbf{y}, \mathbf{x}, \mathbf{w}, \boldsymbol{\mu}, \boldsymbol{\Sigma}, \mathbf{m}, \mathbf{V}, \mathbf{b}, \pi) \\
&= \prod_i p(\mathbf{y}_i|\mathbf{x}_i, \mathbf{w}, b_i) p(\mathbf{x}_i|\boldsymbol{\mu}, \boldsymbol{\Sigma}) p(b_i|\pi) \\
&\quad \times p(\mathbf{w}|\mathbf{m}, \mathbf{V}) p(\boldsymbol{\mu}, \boldsymbol{\Sigma}) p(\mathbf{m}, \mathbf{V}) p(\pi) \quad (10)
\end{align}$$

As it can be seen the likelihood (10) is in an intractable form, since it is not possible to jointly estimate all model parameters directly from (10). That is why the MCMC and VB solutions developed below update every model parameter sequentially from its respective conditional posterior distribution. While the MCMC solution is based on exact conditional posterior distributions, the VB solution utilizes an approximation to the true conditional posterior distribution.

### 4.1. MCMC Update Equations

We derive the MCMC update equations, based on the exact full conditional posteriors of the attack model parameters and construct a Gibbs sampler that iteratively samples from these update equations.

The full conditional posterior distributions of the model parameters are as follows:

- updating $\{\boldsymbol{\mu}, \boldsymbol{\Sigma}\}$.

$$p(\boldsymbol{\mu}, \boldsymbol{\Sigma}|\mathbf{x}) \propto p(\boldsymbol{\mu}, \boldsymbol{\Sigma}) \prod_i p(\mathbf{x}_i|\boldsymbol{\mu}, \boldsymbol{\Sigma})$$
$$\propto \mathcal{IW}(\boldsymbol{\Sigma}|\omega_0 + n, \boldsymbol{\Sigma}_\Sigma) \mathcal{N}(\boldsymbol{\mu}|\boldsymbol{\mu}_\mu, \frac{\boldsymbol{\Sigma}}{n+1}), \quad (11)$$

where

$$\boldsymbol{\Sigma}_\Sigma = \boldsymbol{\Sigma}_0 + \frac{n}{n+1}(\bar{\mathbf{x}} - \boldsymbol{\mu}_0)(\bar{\mathbf{x}} - \boldsymbol{\mu}_0)' + \sum_i (\mathbf{x}_i - \bar{\mathbf{x}})(\mathbf{x}_i - \bar{\mathbf{x}})' \quad (12)$$

$$\boldsymbol{\mu}_\mu = \frac{\boldsymbol{\mu}_0 + n\bar{\mathbf{x}}}{n+1} \quad (13)$$

$$\bar{\mathbf{x}} = \frac{1}{n} \sum_i \mathbf{x}_i. \quad (14)$$

- updating $\mathbf{w}$.

$$p(\mathbf{w}|\mathbf{y}) \propto p(\mathbf{w}|\mathbf{m}, \mathbf{V}) \prod_i 1(b_i = 1) p(\mathbf{y}_i|\boldsymbol{\mu}, \boldsymbol{\Sigma}, \mathbf{w})$$
$$\propto \mathcal{N}(\mathbf{w}|\mathbf{m}_w, \mathbf{V}_w), \quad (15)$$

where

$$\mathbf{V}_w = (\mathbf{V}^{-1} + n_1 \boldsymbol{\Sigma}^{-1})^{-1} \quad (16)$$
$$\mathbf{m}_w = \mathbf{V}_w (\mathbf{V}^{-1}\mathbf{m} + \boldsymbol{\Sigma}^{-1} \sum_i 1(b_i = 1)(\mathbf{y}_i - \boldsymbol{\mu})) \quad (17)$$
$$n_1 = \sum_i 1(b_i = 1), \quad (18)$$

and $1(\cdot)$ is an indicator function.

- updating $b_i$.

$$p(b_i|\hat{\pi}_i) \propto \text{Bernoulli}(b_i|\hat{\pi}_i), \quad (19)$$

where

$$\hat{\pi}_i = \frac{1}{1 + \frac{1-\pi}{\pi} \frac{\mathcal{N}(\mathbf{y}_i|\boldsymbol{\mu}, \boldsymbol{\Sigma})}{\mathcal{N}(\mathbf{y}_i|\boldsymbol{\mu}+\mathbf{m}, \boldsymbol{\Sigma}+\mathbf{V})}}. \quad (20)$$

- updating $\pi$.

$$p(\pi|b) = \text{Beta}(\pi|\hat{a}_\pi, \hat{b}_\pi), \quad (21)$$

where

$$\hat{a}_\pi = a_\pi + \sum_i 1(b_i = 1), \quad (22)$$
$$\hat{b}_\pi = b_\pi + \sum_i 1(b_i = 0). \quad (23)$$

- updating $\{\mathbf{m}, \mathbf{V}\}$.

$$p(\mathbf{m}, \mathbf{V}|\mathbf{w}) \propto p(\mathbf{m}, \mathbf{V}) p(\mathbf{w}|\mathbf{m}, \mathbf{V})$$
$$\propto \mathcal{IW}(\mathbf{V}|\omega_0 + 1, \mathbf{V}_v) \mathcal{N}(\mathbf{m}|\mathbf{m}_m, \frac{\mathbf{V}}{2}), \quad (24)$$

where

$$\mathbf{V}_v = \mathbf{V}_0 + \frac{1}{2}(\mathbf{w} - \mathbf{m}_0)(\mathbf{w} - \mathbf{m}_0)' \quad (25)$$
$$\mathbf{m}_m = \frac{\mathbf{m}_0 + \mathbf{w}}{2}. \quad (26)$$

- updating $\mathbf{x}_i$.

$$\mathbf{x}_i = \begin{cases} \mathbf{y}_i - \mathbf{w} & \text{if } b_i = 1, \\ \mathbf{y}_i & \text{if } b_i = 0 \end{cases} \quad (27)$$

### 4.2. VB Update Equations

The VB approach tries to find a tractable lower bound $\mathcal{L}(q)$ to the logarithm of the marginal likelihood (10), which can be iteratively updated (tightened). If we denote by $\boldsymbol{\theta}$ the model parameters $\{\mathbf{w}, \boldsymbol{\mu}, \boldsymbol{\Sigma}, \mathbf{m}, \mathbf{V}, \mathbf{b}, \pi\}$ that we want to update, the optimal posterior update that gives the tightest bound (Beal & Ghahramani, 2003) is given as

$$q_j(\theta_j) \propto \exp\left(\langle \ln p(\mathbf{y}, \boldsymbol{\theta}) \rangle_{-j}\right), \quad (28)$$

where $\langle \cdot \rangle_{-j}$ denotes expectation with respect to all parameters except for the $j$th parameter that is being updated.

Using (28), the posterior VB updates of the model parameters are as follows:

- updating $\{\boldsymbol{\mu}, \boldsymbol{\Sigma}\}$.

$$q(\boldsymbol{\mu}, \boldsymbol{\Sigma}) \propto \exp\left\langle \ln\left(p(\boldsymbol{\mu}, \boldsymbol{\Sigma}) \prod_i p(\mathbf{x}_i|\boldsymbol{\mu}, \boldsymbol{\Sigma})\right) \right\rangle$$

$$\propto \mathcal{IW}(\boldsymbol{\Sigma}|\omega_0 + n, \boldsymbol{\Sigma}_\Sigma^{vb}) \mathcal{N}\left(\boldsymbol{\mu}|\boldsymbol{\mu}_{\boldsymbol{\mu}}^{vb}, \frac{\langle \boldsymbol{\Sigma}^{-1} \rangle^{-1}}{n+1}\right), \quad (29)$$

where

$$\boldsymbol{\Sigma}_\Sigma^{vb} = \boldsymbol{\Sigma}_0 + \frac{n}{n+1}(\bar{\mathbf{z}} + \boldsymbol{\mu}_0)(\bar{\mathbf{z}} + \boldsymbol{\mu}_0)'$$
$$+ \sum_i \left(\langle b_i^2 \rangle \langle \mathbf{w}\mathbf{w}' \rangle - \langle b_i \rangle^2 \langle \mathbf{w} \rangle \langle \mathbf{w} \rangle'\right)$$
$$+ \sum_i \left(\langle b_i \rangle \langle \mathbf{w} \rangle - \mathbf{y}_i - \bar{\mathbf{z}}\right)\left(\langle b_i \rangle \langle \mathbf{w} \rangle - \mathbf{y}_i - \bar{\mathbf{z}}\right)' \quad (30)$$

$$\boldsymbol{\mu}_{\boldsymbol{\mu}}^{vb} = \frac{\boldsymbol{\mu}_0 - n\bar{\mathbf{z}}}{n+1} \quad (31)$$

$$\bar{\mathbf{z}} = \frac{1}{n}\sum_i \left(\langle b_i \rangle \langle \mathbf{w} \rangle - \mathbf{y}_i\right), \quad (32)$$

and $b_i^2 = b_i$, following the properties of the Bernoulli random variable.

- updating $\mathbf{w}$.

$$q(\mathbf{w}) \propto \exp\left\langle \ln\left(p(\mathbf{w}) \prod_i p(\mathbf{y}_i|\mathbf{w})\right)\right\rangle$$
$$\propto \mathcal{N}(\mathbf{w}|\mathbf{m}_w^{vb}, \mathbf{V}_w^{vb}), \quad (33)$$

where

$$\mathbf{V}_w^{vb} = \left(\langle \mathbf{V}^{-1} \rangle + \langle \boldsymbol{\Sigma}^{-1} \rangle \right.$$
$$\left. \times \sum_i \left(\langle b_i \rangle^2 + \langle b_i - \langle b_i \rangle \rangle^2 \right)\right)^{-1} \quad (34)$$

$$\mathbf{m}_w^{vb} = \mathbf{V}_w^{vb} \left(\langle \mathbf{V}^{-1} \rangle \langle \mathbf{m} \rangle\right.$$
$$\left. + \langle \boldsymbol{\Sigma}^{-1} \rangle \sum_i \langle b_i \rangle (\mathbf{y}_i - \langle \boldsymbol{\mu} \rangle)\right). \quad (35)$$

- updating $b_i$.

$$q(b_i|\hat{\pi}_i) \propto \text{Bernoulli}(b_i|\hat{\pi}_i), \quad (36)$$

where

$$\hat{\pi}_i = \frac{1}{1 + \frac{\exp\langle \ln(1-\pi)\rangle \mathcal{N}\left(\mathbf{y}_i|\langle\boldsymbol{\mu}\rangle, \langle\boldsymbol{\Sigma}^{-1}\rangle^{-1}\right)}{\exp\langle \ln \pi \rangle \mathcal{N}\left(\mathbf{y}_i|\langle\boldsymbol{\mu}\rangle + \langle\mathbf{m}\rangle, \langle\boldsymbol{\Sigma}^{-1}\rangle^{-1} + \langle\mathbf{V}^{-1}\rangle^{-1}\right)}}.$$

- updating $\pi$.

$$q(\pi|b) \propto \text{Beta}(\pi|a_\pi^{vb}, b_\pi^{vb}), \quad (37)$$

where

$$a_\pi^{vb} = a_\pi + \sum_i \langle b_i \rangle \quad (38)$$

$$b_\pi^{vb} = b_\pi + n - \sum_i \langle b_i \rangle. \quad (39)$$

- updating $\{\mathbf{m}, \mathbf{V}\}$.

$$q(\mathbf{m}, \mathbf{V}|\mathbf{w}) \propto \exp\left\langle \ln\left(p(\mathbf{m}, \mathbf{V}) p(\mathbf{w}|\mathbf{m}, \mathbf{V})\right)\right\rangle$$

$$\propto \mathcal{IW}(\mathbf{V}|\omega_0 + 1, \mathbf{V}_v^{vb}) \mathcal{N}\left(\mathbf{m}|\mathbf{m}_m^{vb}, \frac{\langle \mathbf{V}^{-1}\rangle^{-1}}{2}\right), (40)$$

where

$$\mathbf{V}_v^{vb} = \mathbf{V}_0 + \frac{1}{2}(\mathbf{m}_0 - \langle \mathbf{w} \rangle)(\mathbf{m}_0 - \langle \mathbf{w} \rangle)' \quad (41)$$

$$\mathbf{m}_m^{vb} = \frac{\mathbf{m}_0 + \langle \mathbf{w} \rangle}{2}. \quad (42)$$

## 5. Experiments

We perform experiments with both synthetic and real host signals. To quantify the performance of our algorithm, we compute the probability of error $P_e = \frac{1}{n}\sum_i \mathbf{1}(b_i \neq \hat{b}_i)$ and the relative watermark reconstruction error $R_w = \frac{\|\mathbf{w} - \hat{\mathbf{w}}\|_2}{\|\mathbf{w}\|_2}$, where $\|\cdot\|_2$ is the $L_2$ norm, and $\hat{\mathbf{w}}$ is the estimated watermark signal.

In all experiments, the model hyper-parameters are initialized as $a_\pi = b_\pi = 0.5n$, $\boldsymbol{\mu}_0 = \mathbf{m}_0 = \frac{1}{n}\sum_i \mathbf{y}_i$, $\omega_0 = d + 1$, $\boldsymbol{\Sigma}_0 = \frac{1}{n}\sum_i (\mathbf{y}_i - \boldsymbol{\mu}_0)(\mathbf{y}_i - \boldsymbol{\mu}_0)'$ and $\mathbf{V}_0 = \frac{1}{10^{DWR/10}}\boldsymbol{\Sigma}_0$. For each iteration, we computed 95% credible intervals of the individual samples in the watermark signal estimate $\hat{\mathbf{w}}$. With respect to the MCMC solution, we performed 2000 iterations of the Gibbs sampler, discarding the first 1000 as burn in iterations and averaging the results of the remaining 1000 iterations. For the VB solution, we performed 100 iterations and using the last iteration updates of $\langle \mathbf{w} \rangle$, and $\langle b_i \rangle$, as the estimated watermark signal $\hat{\mathbf{w}}$ and message bit $\hat{b}_i$ for $i \in \{1, \ldots, n\}$, respectively.

We implemented the proposed attack model solutions in R, with some of the routines implemented in C/C++. It takes approximately 6 minutes for the MCMC solution to perform 2000 iterations, using 4096 64-dimensional data points. In contrast, the VB solution performs 100 iterations in less than a minute, using the same data points.

### 5.1. Synthetic Host Signals

In this subsection we perform experiments with synthetic host signals. We generated $n = 4096$, $d = 64$-dimensional host data points. Each data point was independent and identically drawn from $\mathcal{N}\left(\mathbf{0}, \mathcal{IW}(d + 1, \mathcal{IW}(d + 1, \mathbf{I}))\right)$, where $\mathbf{I}$ is the identity matrix. In this way, the host signal covariance matrix, although randomly drawn, imposes some structure on the host signal. The watermark signal was drawn from a multivariate Gaussian with mean $\mathcal{N}\left(\mathbf{0}, \mathcal{IW}(d + 1, \mathbf{I})\right)$ and covariance $\mathcal{IW}(d+1, \mathbf{I})$. The watermark signal was zero mean transformed and scaled so that DWR = 30db. The watermark message bits were drawn from Bernoulli(0.5), and the watermarked signal was formed by additive spread-spectrum modulation. The host image and the difference between the watermarked and host images are shown in Fig. 4. Each block of pixels was formed by row-wise transformation of the data point into an $8 \times 8$ matrix. The blocks were then ordered row-wise to form the whole image.

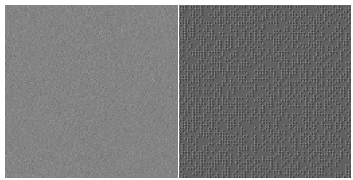

Figure 4. Synthetic host image (left) and watermark signal modulated by the message bits (right). DWR = 30db.

Experimental results of the difference $b_i - \hat{b}_i$ for $i \in \{1, \ldots, n\}$, the watermark signal $\mathbf{w}$ and its estimate $\hat{\mathbf{w}}$ for the MCMC and VB solutions are shown in Fig. 5. The results show that the algorithm was able to obtain a good estimate of the watermark signal and message bitstream, with only a small fraction of misidentified bits. Based on the experimental results, we can see that the MCMC and VB solutions perform comparably in terms of $P_e$ and $R_w$.

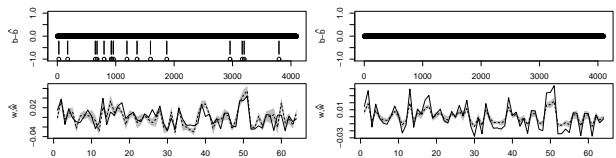

Figure 5. Experimental results for the MCMC (left column) and VB (right column) solutions applied to the synthetic host signal. The plots present $b_i - \hat{b}_i$, the watermark signal $\mathbf{w}$ (solid line) and its estimate $\hat{\mathbf{w}}$ (dashed line). The gray regions represent 95% credible intervals. For the MCMC solution $P_e = 0.004$ and $R_w = 0.48$. For the VB solution $P_e = 0$ and $R_w = 0.256$. Chosen DWR = 30db.

### 5.2. Real Host Signals

In this subsection we perform experiments with real image signals. We applied our algorithm on gray scale images. In the experiments we used image sizes of $512 \times 512$ and $1024 \times 1024$ pixels. The images were split in $8 \times 8$ patches of pixels, making a total of $n = 4096$ and $n = 16384$ patches respectively. The pixels within each patch were concatenated row-wise to form the $d$-dimensional ($d = 64$) data points. The host signal was then normalized to have zero mean.

As in the case of synthetic host signals, the watermark signal was drawn from a multivariate Gaussian with mean $\mathcal{N}(\mathbf{0}, \mathcal{IW}(d+1, \mathbf{I}))$ and covariance $\mathcal{IW}(d+1, \mathbf{I})$. The watermark signal was then scaled such that DWR = 30db and embedded by additive spread-spectrum modulation with $\Pr(b_i = 1) = 0.5$, for $i \in \{1 \ldots n\}$. The real host images used in the experiments are shown in Fig. 6.

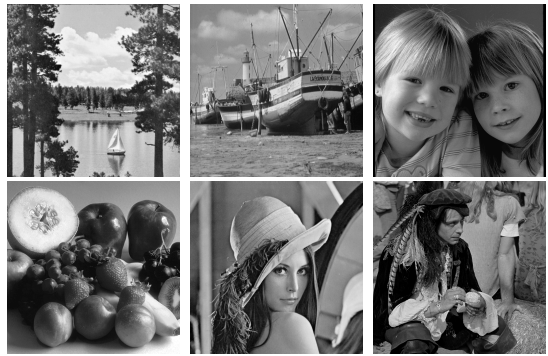

Figure 6. Real host images. From top left to bottom right: *Lake, Boat, Children, Fruits, Lena, Pirate.*

Experimental results for the MCMC and VB solutions applied to the real host images from Fig. 6 are shown in Fig. 7. It can be seen that both solutions perform comparably with respect to real host signals.

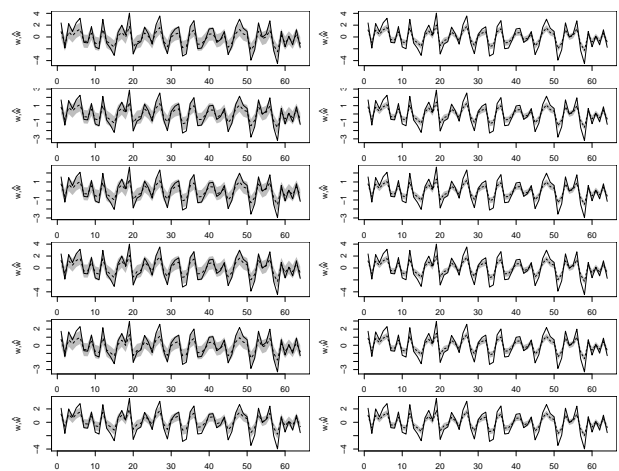

Figure 7. Experimental results for the MCMC (left column) and VB (right column) solutions. The solid line represents the watermark signal $\mathbf{w}$ and the dashed line is its estimate $\hat{\mathbf{w}}$. The gray regions represent 95% credible intervals. From top to bottom: *Lake, Boat, Children, Fruits, Lena, Pirate.* Chosen DWR = 30db.

Computations of the lower bound $\mathcal{L}(q)$ for the VB solution applied on the real host images in Fig. 6 are shown in Fig. 8. The results show that the VB so-lution converges in less than 20 iterations for all real host images considered in the experiments.

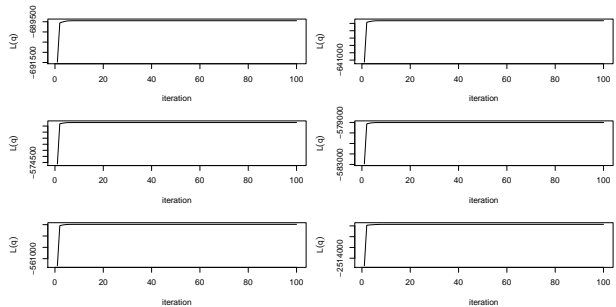

*Figure 8.* Experimental results for the lower bound to the log-likelihood. Please see supplementary for a detailed expression of $\mathcal{L}(q)$. From top left to bottom right: *Lake, Boat, Children, Fruits, Lena, Pirate.* Chosen DWR = 30db.

To quantify the performance of the attack model with respect to different DWR levels, we perform experiments with the real host images in Fig. 6 and varying the DWR $\in \{20, \ldots, 40\}$. The interval of values for the DWR was chosen so that the middle is at DWR = 30db, at which level no perceptual degradation to the host image was observed. The watermark signal was drawn from a multivariate Gaussian with mean $\mathcal{N}(\mathbf{0}, \mathcal{IW}(d+1, \mathbf{I}))$ and covariance $\mathcal{IW}(d+1, \mathbf{I})$. For each host image, the watermark signal was drawn only once and then zero mean transformed, and scaled down differently to achieve the different DWR levels. Experimental results of $P_e$ and $R_w$ as functions of DWR are shown in Fig. 9 and Fig. 10 respectively, with both solutions performing comparably.

## 6. Discussion

We presented a new attack model on repetitive spread-spectrum watermarking systems, based on Bayesian statistics. The proposed algorithm jointly estimates the watermark signal and message bitstream, directly from the watermarked signal and without access to the watermark decoder. We developed MCMC and VB solutions that perform comparably in terms of probability of error and relative watermark reconstruction error, on both synthetic and real host signals. Fast convergence is observed in both solutions, particularly in the VB solution where the algorithm converges in less than 20 iterations, with both synthetic and real host signals. While the MCMC solution is expected to result in more accurate estimates for infinite number of iterations, the VB solution is computationally more efficient and therefore more appropriate for large data sets. We demonstrated that the attack model is able to correctly infer a large part of the message bitstream while at the same time obtaining a good estimate of the watermark signal.

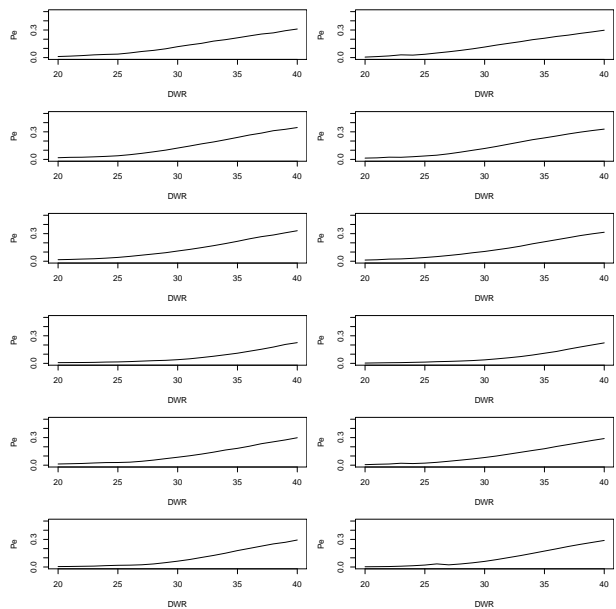

*Figure 9.* Experimental results of $P_e$, based on the MCMC (left column) and VB (right column) solutions. From top to bottom: *Lake, Boat, Children, Fruits, Lena, Pirate.*

## 7. Acknowledgments

The authors wish to acknowledge the helpful suggestions of the reviewers.

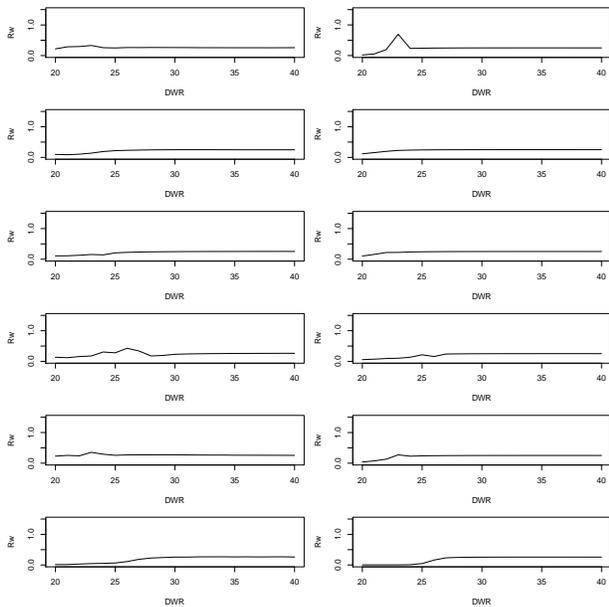

*Figure 10.* Experimental results of $R_w$, based on MCMC (left column) and VB (right column) solutions. From top to bottom: *Lake, Boat, Children, Fruits, Lena, Pirate*.